\documentclass{elsart}

\usepackage{graphics}
\usepackage{amsmath}
\usepackage[dvips]{graphicx}
\usepackage{amsfonts,amsbsy}
\usepackage{amssymb}
\usepackage{bbm}
\newcommand{\beq}{\begin{equation}}
\newcommand{\eeq}{\end{equation}}
\newcommand{\beqa}{\begin{eqnarray}}
\newcommand{\eeqa}{\end{eqnarray}}
\def\r{{\boldsymbol r}}
\def\s{{\boldsymbol s}}
\def\b{{\boldsymbol b}}
\def\d{{\boldsymbol d}}
\def\z{{\boldsymbol z}}
\def\x{{\boldsymbol x}}

\def\y{{\boldsymbol y}}
\def\k{{\boldsymbol k}}

\def\p{{\boldsymbol p}}
\def\0{{\boldsymbol 0}}

\def\E{{\boldsymbol E}}

\def\cal{\mathcal}
\newcommand{\del}{\partial}

\begin{document}
\begin{frontmatter}

\title{The Status of Parton Saturation and the CGC}
\author{J.P. Blaizot$^1$}
\address{$^1$ IPhT, CEA Saclay, 91191 Gif-sur-Yvette cedex, France}

\date{\today}

\begin{abstract} 
This is a personal summary of the meeting  ``Saturation, the Color Glass Condensate and Glasma:
What Have we Learned from RHIC?''  that took place at BNL in May 2010. The purpose of the meeting was to discuss the status of  high density QCD and parton saturation, and to review the progress that  RHIC has allowed in the field.\end{abstract}
%


\end{frontmatter}
%
\section{Introduction}%

Understanding nucleus-nucleus collisions at high energies requires a good knowledge of that part of the nuclear wave functions that describes ``low $x$''  degrees of freedom. Let us recall indeed that most particles produced in such collisions have momenta in the GeV range or less, and the partons involved in the production process have momentum fraction $x=({m_\perp}/{\sqrt{s}}){\rm e}^{\pm y}$ where $\sqrt {s}$ is the nucleon-nucleon center of mass energy and $y$ the rapidity of the produced particle. Taking $m_\perp$ to be $\sim$1 GeV, one sees that particles produced at mid rapidity ($y\simeq 0$) at RHIC probe the nuclear wave functions at values of $x\lesssim 10^{-2}$, and at much smaller values of $x$  at forward rapidity. At the full LHC energy, values as small as $10^{-4}$ will be reached at mid rapidity. These values are comparable to the values probed at HERA in deep inelastic collisions on protons.  

It is well established that in this regime of small $x$, the gluons dominate the hadron wave functions. The gluon density is large and grows with lowering $x$, but this growth eventually saturates: one reaches then the regime of `parton saturation', which has been much studied over the last decade. Saturation leads to a simple structure in the plane  ($\ln (1/x), \ln Q^2$), whose coordinate axis are, loosely speaking,  the longitudinal and the transverse components of the parton momenta: a  line separates dense and dilute parton systems. This line is parameterized by a `saturation momentum' $Q_s$, a growing function of $1/x$: partons with transverse momentum $k_\perp\gg Q_s$ are in the dilute regime, those with $k_\perp \lesssim Q_s$ are in the dense saturated regime. The determination of the saturation momentum  requires the solution of non linear evolution equations that incorporate physics beyond that of the linear evolution equations commonly used to fit deep inelastic scattering data. In the vicinity of saturation, perturbation theory breaks down: the large gluon density compensates for the weakness of the coupling, making the effective expansion parameter of order unity. 

While the gross features of the saturation regime are well identified, the physics of saturation, or more generally of dense partonic systems, is a
multifaceted many-body problem that has been approached using various formalisms, each one emphasizing one particular aspect of the problem. These formalisms often become rapidly very technical, which makes it difficult to compare their physical contents, and relate their predictions to experimental data. In view of this, I have chosen in this presentation to put the various facets of parton saturation, as they emerge from different formalisms, in a balanced perspective. I shall keep the discussion  simple, my goal being to reveal as much as possible the connections between the various points of view in physical terms, rather than dwelling on their technical intricacies.  I hope the exercise will have some usefulness.

There are excellent reviews on the formal and phenomenological developments to which I  refer the reader for complete references  \cite{Iancu:2003xm,Weigert:2005us,Triantafyllopoulos:2005cn,JalilianMarian:2005jf,Gelis:2010nm,Lappi:2010ek,Frankfurt:2005mc,Munier:2009pc} (see also the relevant lectures in \cite{Blaizot:2002sr}). I shall indeed not systematically quote the original literature, except in cases where I am using explicitly a specific result.  Contributions to the workshop will be referred to  by the names of their authors. I should however mention that, because I have chosen to put emphasis on the conceptual issues rather than going through a systematic account of the presentations, I shall not be able to cover all the talks that were presented.  
 
This paper is organized as follows. The next three sections introduce the physics of parton saturation from three different perspectives.  First I briefly review the familiar  evolution of parton distributions as a function of $x$ and $Q^2$, as describred by linear equations that are solidly rooted in standard techniques of perturbative QCD. I also discuss there the early description of parton saturation as a balance between gluon splitting and gluon recombination.  In the following section, I introduce further concepts, that grew from  the analysis of deep inelastic lepton-hadron collisions, namely the eikonal approximation, Wilson lines and color dipoles. The study of the propagation of a color dipole through the color field of a hadron will provide another point of view on saturation, which is seen there as arising from the  multiple scatterings on the dense system of gluons. Finally, I turn to the Color Glass Condensate  (CGC) where the classical fields are given  a prominent role. The CGC aims at a complete description of the small $x$ part of hadron wave functions that can be used to calculate many processes dominated by small $x$ partons. The phenomenological applications are presented in the next to last section, while the  last section summarizes the conclusions.

\section{The wave function of a hadron at high energy}

At the beginning of this discussion it is perhaps appropriate to recall that the notion of a wave-function for a hadron at high energy suffers from well-known ambiguities: it depends on the frame where it is defined, on the gauge chosen,  with the parton picture emerging more naturally in the light-cone gauge and in the infinite momentum frame. A further ambiguity arises often in higher order calculations of a given process in the separation of the constituents of the hadrons from the probe that is used to measure them. 

The wave function of a hadron is commonly  characterized in terms of `partons' carrying momentum ${\k}=(k_z,\k_\perp)$, with the longitudinal momentum $k_z$ given as a fraction $x$ of the momentum of the parent hadron, $k_z=x P_z$. We shall often in the following refer to  light cone coordinates, with $x^\pm=(t\pm z)/\sqrt{2}$. In these coordinates, a right mover parton has momentum $k^+= x P^+$. 
 
\begin{figure}[htbp]
\begin{center}
\vspace{0.5cm}
\includegraphics[scale=0.5]{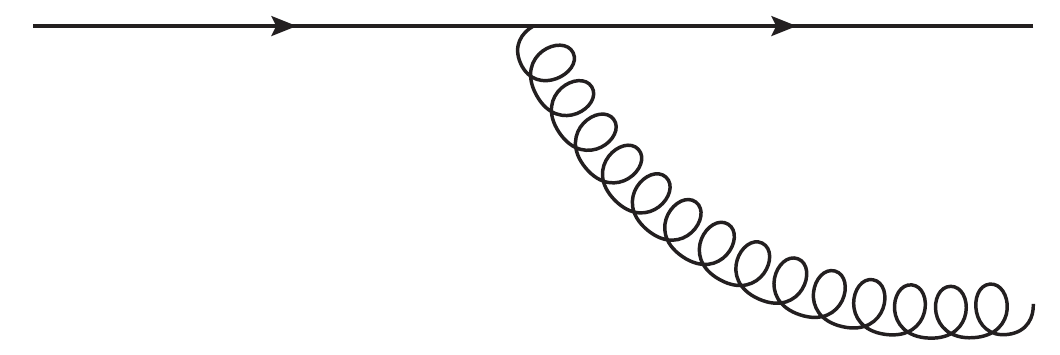}
\end{center}
\caption{\label{fig:brem} The elementary branching process: a parton (e.g a valence quark represented here by the straight line) emits a gluon (the wavy line) with momentum $k_z=(xP_z, \k_\perp)$, with $P_z$ the momentum of the parent hadron to which the initial parton belongs.}
\end{figure}
The basic phenomenon that controls the evolution of the wave functions is the branching of partons, with the elementary process displayed in Fig.~\ref{fig:brem}. This process, which corresponds to the radiation of a soft gluon from either a quark or a gluon, occurs with a probability
\beq\label{eq:branching}
dP\simeq \frac{\alpha_s C_R}{\pi^2}\,\frac{d^2 k_\perp}{k_\perp^2}\,\frac{dx}{x},
\eeq
with $C_R=C_A=N_c$ for the radiation from a gluon, and $C_R=C_F=(N_c^2-1)/2N_c$ for a radiation from a quark. As revealed by Eq.~(\ref{eq:branching}), this probability is enhanced when the emitted gluon carries a small transverse momentum or a small energy fraction. 

The state of a hadron is
built up from successive splittings of partons starting from the valence quarks.
 For just a valence quark, leading order perturbation theory yields the integrated gluon distribution $xG(x,Q^2)$ as
\beq\label{Gvalencequark}
xG(x,Q^2)=\frac{\alpha_s C_F}{\pi}\ln\left(  \frac{Q^2}{\Lambda_{QCD}^2} \right).
\eeq
Roughly speaking,  $xG(x,Q^2)$ counts the number of gluons in the hadron wave function (here the valence quark) with longitudinal momentum between $xP^+ $ and $(x+dx)P^+$, and localized in the transverse plane to a region of size $\Delta x_\perp\sim 1/Q$ (corresponding to the wavelength of the partons with the highest transverse momenta). The infrared cut-off $\sim\Lambda_{QCD}$ accounts for the fact that the parton description ceases to make sense for partons that have wavelengths larger than the typical confinement scale $r_0\sim 1/\Lambda_{QCD}$.

It is convenient to relate  $xG(x,Q^2)$ to a so-called ``unintegrated parton distribution'' $\varphi(x,k_\perp)$ more directly related to the phase space density. We write
\beq\label{xGsurR2}
\frac{xG(x,Q^2)}{\pi R^2}=\int^{Q} d^2\k_\perp \frac{dN}{dy d^2\k_\perp d^2\b},
\eeq with $dy=dx/x$ and 
\beq
\frac{dN}{dy d^2\k_\perp d^2\b}=\frac{2(N_c^2-1)}{(2\pi)^3}\, \varphi(x,\k_\perp).
\eeq
As is clear from its definition $\varphi(x,\k_\perp)$ denotes the density in the transverse plane  of gluons with transverse momentum $\k_\perp$ and a definite spin and color.  

The density of partons in the cloud of gluons that surrounds a valence quark is not a fixed quantity but a quantity that depends on the resolution with which one is probing the gluon cloud, as one can already see from Eq.~(\ref{Gvalencequark}). If one increases the resolution, by increasing $Q^2$, on sees more and more partons, i.e., the parton density increases. The formula (\ref{Gvalencequark}) has been obtained by using leading order perturbation theory, that is, by taking into account a single branching. But even if $\alpha_s$ is  small, successive branchings must be taken into account when $\ln Q^2$ becomes large: a large value of $Q^2$ is primarily  achieved through successive emissions of partons with moderate transverse momenta. In technical terms, when $\alpha_s\ln Q^2$ becomes of order unity, higher order terms become significant and must be resummed. This resummation is achieved by the DGLAP equation \cite{DGLAP}, which we write schematically as\beq
Q^2\frac{\del }{\del Q^2}G(x,Q^2)=\frac{\alpha_s(Q^2)}{2\pi}\int_x^1\frac{dz}{z}P(x/z)G(z,Q^2),
\eeq
where $P(x/z)$, the `splitting function',  gives the probability that a  daughter parton with momentum $x$ is produced by the splitting of a parent parton with momentum $z$.

The DGLAP equation leads to an increase of the parton density with increasing $Q^2$.  However, this increase is slow, involving typically $\ln Q^2$. Since the size of the added partons decreases as $1/Q^2$, the area occupied by these new partons  in the transverse plane eventually decreases with increasing $Q^2$. Thus, even though the density increases, the system of partons produced by the DGLAP evolution is effectively more and more dilute with the partons effectively weakly coupled. As $Q^2$ grows (with $x$ kept not too small) perturbation theory becomes more and more reliable in describing the changes in the hadron wave functions. 

When one increases the rapidity (or equivalently the energy), $1/x$ decreases and one eventually reaches a regime where new corrections become important: when $\alpha_s\ln (1/x)$ becomes of order unity, the corresponding large logarithms need to be resummed. This new resummation is achieved by the BFKL equation \cite{BFKL}. Before we turn to this equation, let us observe that there is a regime where both large $Q^2$ and small $x$ effects are simultaneously taken into account, albeit partially. This is the so-called Double Logarithmic Limit (DLL), where $\alpha_s \ln(1/x)\ln Q^2\simeq 1$. The evolution equation in this limit (and at fixed coupling) reads
\beq\label{DLL}
\frac{\del^2}{\del\ln(1/x)\del \ln Q^2} \,xG(x,Q^2)=\frac{\alpha_s C_A}{\pi} \, xG(x,Q^2),
\eeq
whose solution is 
$
xG(x,Q^2)\propto \exp\left\{   2\sqrt{\bar \alpha_s \ln\frac{1}{x}\ln\frac{Q^2}{Q_0^2}}  \right\},
$ with $\bar\alpha_s\equiv \alpha_s C_A/\pi$.
This solution reveals the growth of the structure function at small $x$, a  growth which is however milder than that predicted by the full BFKL equation to which we now turn. 

Written as an equation for the unintegrated gluon density, the  BFKL equation takes the form ($y=\ln(1/x)$)
\beq
\frac{\del \varphi(y,\k_\perp)}{\del y}=\bar\alpha_s \int\frac{d^2\p_\perp}{\pi}\,\frac{\k_\perp^2}{\p_\perp^2(\k_\perp-\p_\perp)^2}\left[  \varphi(y,\p_\perp)-\frac{1}{2} \varphi(y,\k_\perp)  \right].
\eeq
As it should, it contains the DLL   obtained when $k^2$ is sufficiently large. But the most remarkable feature of the BFKL evolution is the exponential growth that it predicts for the gluon density:
\beq
\varphi(y,\k_\perp^2)\sim {\rm e}^{\omega\bar\alpha_s y},
\eeq
with $\omega=4\ln 2$ (in leading order). This explosive growth may be traced back to important properties of the BFKL cascade: the evolution takes place at (approximately) fixed transverse momentum, and the longitudinal momenta for the successive gluon emissions  $i$ and $i+1$ are ordered, namely  $x_{i+1}<x_i$. Since the typical lifetime of a fluctuation is $\Delta t_i\sim 2x_iP^+/k_\perp^2$,  $x_{i+1}<x_i$ implies $\Delta t_i>\Delta t_{i+1}$. Thus,  the $n$th gluon is effectively emitted from a strong color source, made from the gluons emitted at $x_{i<n}$, which overlap in the transverse plane and which can be considered as frozen during this last emission ($\Delta t_n<\Delta t_{i<n}$). This enhanced color charge triggers further emission leading to a kind of a chain reaction responsible for the exponential increase \cite{Mueller:1994up}.

It was recognized early on that this growth of the gluon density, predicted by the linear BFKL equation, could not go on for ever,  and various mechanisms leading to a `saturation' of the process have been looked for. 
The early approaches to saturation invoked a non linear contribution to the evolution equation \cite{Gribov:1984tu,Mueller:1985wy} and leads (schematically) to an equation of the form
\beq
\frac{\partial^2\; xG(x,Q^2)}{\partial \ln(1/x) \,\partial \ln Q^2}=\bar\alpha_s \,xG(x,Q^2)-\frac{9}{16}\bar\alpha_s^2\,\pi^2\frac{[xG(x,Q^2)]^2}{R^2Q^2} 
\eeq
which differs from  Eq.~(\ref{DLL}) by the second term accounting for `gluon recombination'. (The square of the gluon distribution is an approximation for a 2-point gluon distribution, in the form used in \cite{Blaizot:1987nc}.) This ``kinetic'' vision of gluon saturation suggests immediately the existence of a characteristic momentum scale at which the processes of gluon emission and gluon recombination balance each other. This saturation momentum $Q_s$ is (parametrically) given by
\begin{equation}\label{Qsaturation}
 Q_s^2
\sim \alpha_s(Q_s^2)\frac{xG(x,Q_s^2)}{\pi R^2}\; .
\end{equation}
Referring to Eq.~(\ref{xGsurR2}), one sees that at saturation the phase space density of modes with $k^2_\perp\lesssim Q^2$ is large,  of order $1/\alpha_s$. This is an important feature of saturation, which we shall often refer to.
\begin{figure}[htbp]
\begin{center}
\includegraphics[scale=0.8]{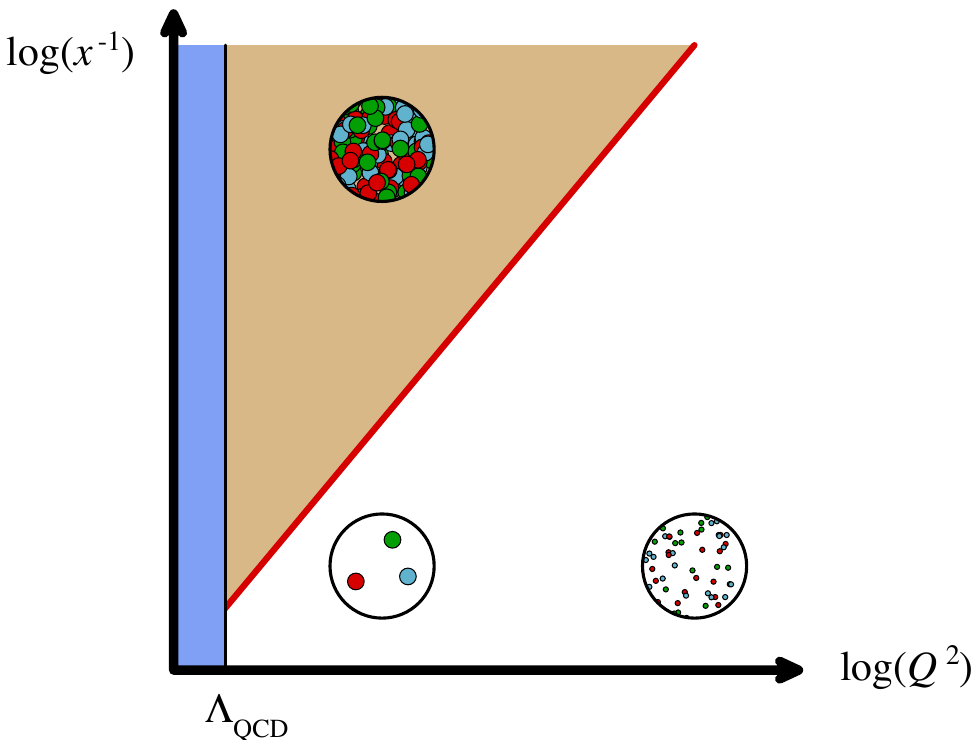}
\end{center}
\caption{\label{fig:xQplane} The various regimes in the plane $\ln (1/x), \ln Q^2$ (from F. Gelis). At large $Q^2$, and moderate values of $x$, partons are weakly coupled and form a dilute system. At small values of $x$, partons form a dense (`saturated') system and are strongly coupled (shaded area). The region with $Q^2<\Lambda_{QCD}^2$ is genuinely non perturbative and not amenable to a weak coupling description.}
\vspace{0.5cm}
\end{figure}

In slightly more general terms,  the onset of saturation coincides with a breakdown of perturbation theory where 
gluon interaction energies become comparable to their transverse  kinetic energies, that is $\partial ^2\sim \alpha_s
\langle A^2\rangle_Q$, where  $\langle A^2 \rangle_Q$ denotes the
fluctuations of the gauge fields with transverse momenta up to $Q$, $\langle A^2 \rangle_Q\sim xG(x,Q^2)/\pi R^2$. Thus, as illustrated in Fig.~\ref{fig:xQplane}, the saturation momentum separates partons into dilute modes with $k_\perp\gg Q_s$ that are weakly coupled, and modes with $k_\perp\lesssim Q_s$ that are strongly coupled because densely populated.

 \section{Color dipoles, Wilson lines}

\begin{figure}[htbp]
\begin{center}
\includegraphics[scale=0.4]{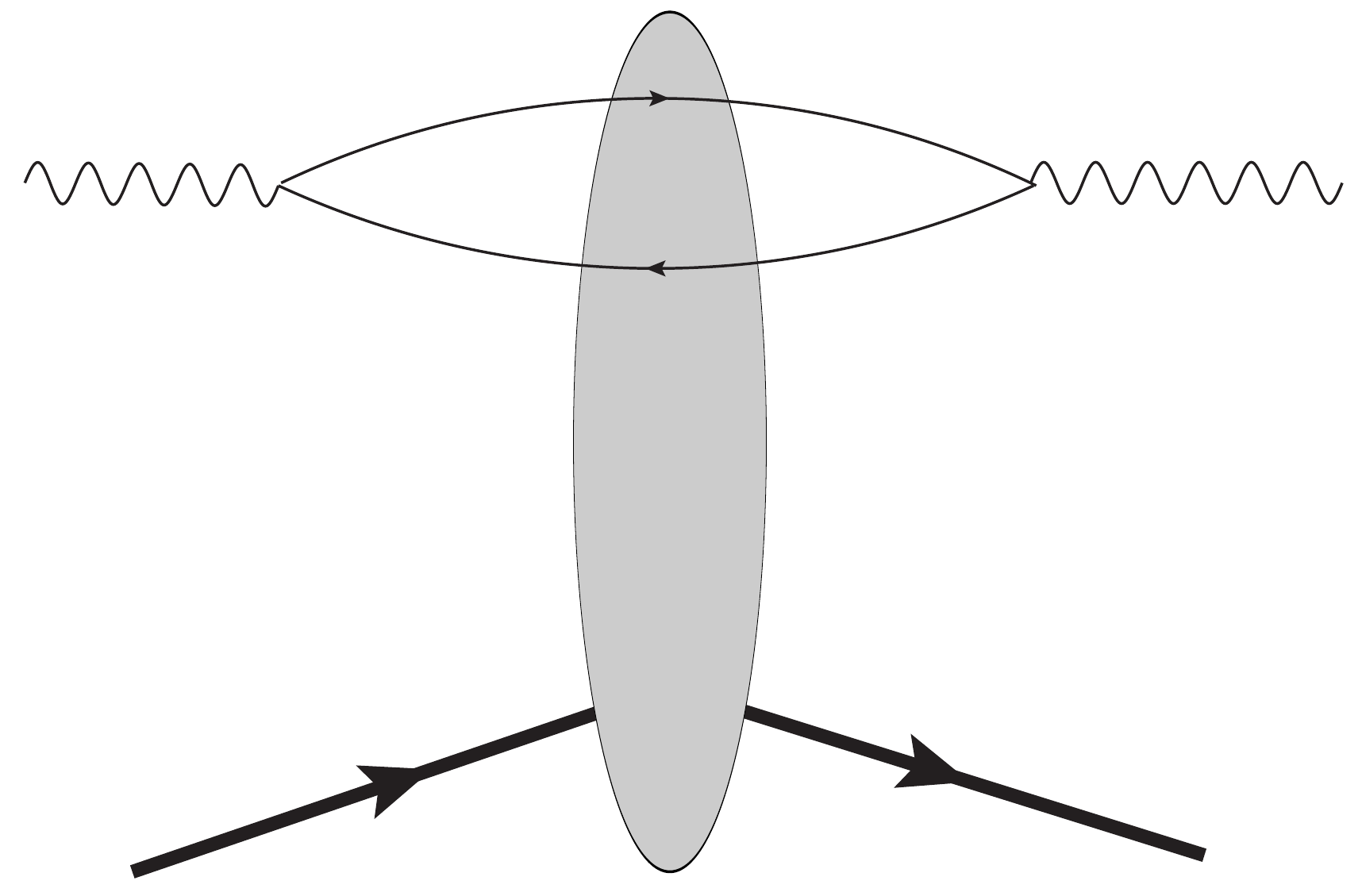}
\end{center}
\caption{\label{fig:dipole} Pictorial view of the interaction of a virtual photon and a hadron. In the chosen frame, the virtual photon splits into a $q\bar q$ pair, a color dipole, that subsequently interacts with the fluctuating color field of the hadron. At high energy, the propagation of the dipole through the hadron can be treated in the eikonal approximation, i.e., by Wilson lines, and during the passage of the dipole through the hadron, it ``sees'' a frozen color field $A^+$, schematically indicated by the shaded area.}
\end{figure}

Another view of saturation, that will eventually lead us to the most elaborate non linear evolution equations, builds on a picture commonly used in the analysis of lepton-hadron deep-inelastic scattering (DIS). In an appropriate frame, one can describe  the interaction of the virtual photon with the hadron  as the interaction of a color $q\bar q$ dipole (emerging form the photon) with the color field of the hadron (see Fig.~\ref{fig:dipole}).
The  factor in the interaction cross section that is relevant for the present discussion is 
$\sigma_{\rm dip}(x,\r_\perp)$,  the total dipole-hadron 
cross-section (that we assume for simplicity here to be a function of $x$ and $\r_\perp$ -- for a recent and thorough analysis of this picture see \cite{Ewerz:2004vf}).
This dipole cross section  can be calculated in the eikonal approximation, with the size $\r_\perp$ of the dipole remaining unchanged during the interaction. In this approximation, the S-matrix for the scattering of a quark moving in the negative $z$ direction is  given by the Wilson line
\begin{equation}
U(\x_\perp)\equiv {\cal P} \exp\left[ -ig \int_{-\infty}^{+\infty} dz^-
A^+(z^-,\x_\perp)\right]\; ,
\end{equation}
where ${\cal P}$ denotes an ordering along the $x^-$ axis, and $A^+$ is
the classical (frozen) color field of the  hadron moving close to the speed
of light in the $+z$ direction. The S-matrix for the scattering of the dipole  contains another, complex conjugate, Wilson line. Keeping in mind that after averaging over the field of the hadron the $S$-matrix will be real, we can write the total dipole cross section as $\sigma_{dip}=2\int d^2\b\, (1-S(\b,\r_\perp))$
\begin{equation}
S(\b,\r_\perp)=\frac{1}{N_c}
{\rm Tr} \left<U(\b+\frac{\r_\perp}{2})\,
U^\dagger(\b-\frac{\r_\perp}{2})\right> .
\label{eq:dipole-cs}
\end{equation}
In fact  we shall ignore here the impact parameter dependence and write simply $S(\r_\perp)$. It is also customary to set $S=1-N$, with $N$ denoting the imaginary part of the forward scattering amplitude. (The real part  is negligible at high energy, and would disappear anyway in the gaussian averages that we are going to perform.) Obviously, when $r_\perp\to 0$, the scattering amplitude vanishes, independently of the field configuration, a characteristic property of the dipole interaction often referred to as `color transparency'. 

As a first orientation on the effect of the averaging over the field configurations (denoted by $\langle \cdots\rangle$ in Eq.~(\ref{eq:dipole-cs})), we may assume that their distribution is gaussian. Then Wick's theorem leads immediately to the following expression
\beq\label{correlatorWL}
S(\r_\perp)=\exp\left\{-g^2C_F \int \frac{d^2\k_\perp}{(2\pi)^2}\frac{  1-{\rm e}^{i\k_\perp\cdot\r_\perp}}{\k_\perp^4} \mu_A^2(\k_\perp)  \right\}, 
\eeq
which involves the correlator  of the color charges of the hadron: $\langle \rho^a(\k_\perp)\rho^b(-\k_\perp)\rangle=\delta_{ab}\pi R^2  \mu_A^2(\k_\perp)$. Specifically, 
we have assumed, for the correlator of two fields (after integrating over the longitudinal coordinate $x^-$, $A^+_a(\x_\perp)\equiv\int dx^- A^+_a(x^-,\x_\perp)$, and assuming that fields at different  $x^-$ are not correlated), 
\beq\label{correlatorA+}
\langle A^+_a(\x_\perp) A^+_b(\y_\perp)\rangle=\delta_{ab}\int \frac{d^2\k_\perp}{(2\pi)^2}\frac{  1-{\rm e}^{i\k_\perp\cdot\r_\perp}}{\k_\perp^4}  \mu_A^2(\k_\perp),
\eeq
with $\r_\perp=\x_\perp-\y_\perp$. 
In the MV model, to be discussed below, all correlations between the color charges in the transverse plane are ignored. In that case $\mu_A^2(\k)=\mu_A^2$ is a constant, $\mu_A^2=N^{val}\, \mu^2$, with $\mu^2$ corresponding to a single valence quark, for which 
$\langle \rho_a(\k_\perp)\rho_b(-\k_\perp)\rangle =\delta_{ab} {g^2}/{2 N_c}$,  so that $\mu^2={g^2}/(2\pi R^2 N_c)
$.

The momentum integral in Eq.~(\ref{correlatorWL}) is reminiscent of Coulomb scattering over a distribution of charges, with the second term accounting for the interference responsible for color transparency. In fact the integral is very well approximated by its contribution for $k_\perp\lesssim 1/r_\perp$,
\beq
\int \frac{d^2\k_\perp}{(2\pi)^2}\frac{  1-{\rm e}^{i\k_\perp\cdot\r_\perp}}{\k_\perp^4}  \approx \frac{r_\perp^2}{16\pi}\ln\frac{r_0^2}{r_\perp^2}.
\eeq
The infrared cut-off $r_0\sim 1/\Lambda_{QCD}$ has been introduced to account for the screening of  the Coulomb field over such distance scales. (We assume $r_\perp\ll r_0$ throughout.) The factor $\r_\perp^2$ in the equation above is characteristic of dipole interaction. The total cross section will depend on the ratio between $r_\perp$ and another length scale, call it $r_s$, that is determined entirely by the dynamics of the field with which the dipole interact. More concretely, we may write, 
\beq\label{dipolesat}
S(\r_\perp)={\rm e}^{-Q_s^2 r_\perp^2/4}
\eeq
with $Q_s=1/r_s$ (the factor 4 is conventional). In the case where the field is created by a random distribution of  ($N^{val}=AN_c$) valence quarks, we have
\beq
Q_s^2=\tilde Q_s^2\ln\frac{r_0^2}{r_\perp^2}= \pi \mu_A^2 xG(x,1/r_\perp^2),\qquad \tilde Q_s^2=\alpha_s C_F\mu_A^2,
\eeq
where we have absorbed the logarithm into the gluon density of a single valence quark at the scale $Q^2\sim 1/r_\perp^2$ (see Eq.~(\ref{Gvalencequark})). Equation (\ref{dipolesat}) exhibits the expected change of regime in the interaction of the dipole with the field of the hadron. A small dipole, with $r_\perp\ll r_s$, is little affected, and its scattering amplitude measures directly the gluon density. A large dipole ($r_\perp\gg r_s$) on the other hand is strongly absorbed: its cross section saturates to the black disk limit, and it is not capable to resolve the parton structure of the field in which it propagates. 

It is in fact interesting to push this discussion a bit further, and note that the inelastic cross section can be written as 
$\sigma_{inel}=\int d^2\b \,(1-S^2(\b,\r_\perp))$, with  
\beq\label{survival}
S^2(\r_\perp)={\rm e}^{-\frac{Q_s^2 \r_\perp^2}{2} }={\rm e}^{-\sigma_{0} (\r_\perp) n(b)}={\rm e}^{-L/\lambda},
\eeq
where $n(b)$ is the density of valence quarks per unit transverse area at impact parameter $b$ (we take it independent of $b$, $n(b)=N^{val}/\pi R^2$), and the last equality expresses $S^2$  as a survival probability with $\lambda$ a mean free path, and $L $ the path length. The dipole cross section, corresponding to the scattering off a single valence quark, is given by 
\beq\label{2gluonexch}
\sigma_{0}(\r_\perp)= \frac{\pi^2 \alpha_s}{N_c} r_\perp^2 xG(x,1/r_\perp^2).
\eeq
So what the eikonal approximation does, when coupled to the gaussian averaging over the field fluctuations, is to express  the survival probability as the  exponential (\ref{survival}), where the cross section $\sigma_0$ is the  two gluon exchange estimate (\ref{2gluonexch}) of the dipole cross section. This allows one to view saturation as resulting from  the multiple scatterings that the dipole undergoes as it traverses the hadron. Referring back to the origin of the calculation in terms of Wilson lines, one sees that when multiple scatterings are important one can no longer expand the Wilson lines: The weak field approximation breaks down. We are in a regime of strong field, an aspect of saturation that we have already encountered. 

It may be useful to note that much of the previous discussion would apply as well to the scattering of an electric dipole $\d$  on a collection of random charges. The Hamiltonian of this system would read $H=-\d\cdot\E$ and the associated survival probability would be given by
\beq\label{elecdipole}
S^2={\rm e}^{-\frac{1}{2} {\d^2}/{r_s^2}},\qquad   \frac{1}{r_s^2}=\langle \E^2 T^2\rangle
\eeq
where $T$ denotes the time the dipole is exposed to the field ($T$  plays the role of the length of the $x^-$ integration).  In each passage of the dipole through the field, the field is constant so that the wave function of the dipole is just modified by a phase (I assume that $\d$ and $\E$ are parallel for simplicity). But if we take the field to be a random variable, and take the distribution of this variable to be a gaussian, we arrive at Eq. ~(\ref{elecdipole}). To make the connection with the previous calculation involving color fields, we note  that the fluctuation of the color electric field $E_a^i=F^{+i}_a=\nabla_i A^+$ can be obtained from the gluon distribution  (again for the field created by a  random distribution of valence quarks) as (using Eq.~(\ref{correlatorA+}))
\beq
\langle \E(\x_\perp)\cdot \E(\x_\perp)\rangle= \frac{1}{R^2} xG(x,Q^2), 
\eeq
 so that $\sigma_0$ can indeed be written as
\beq
n(b)\sigma_0=\frac{1}{2} \, \frac{g^2}{2N_c}\, \langle  \E(\x_\perp)\cdot \E(\x_\perp)\rangle\, r_\perp^2.
\eeq

This picture of the dipole as a probe of the hadron wave function leads one to expect a rather simple property that will be more sharply defined later. We note that all the dependence on the wave function is contained in the scale $r_s$ that characterizes the change of regime between the dilute regime and the saturated one. The  energy dependence of $r_s$ is determined by the dynamics of gauge fields, but we expect $r_s$ to appear in the cross section only in the ratio $r^2_\perp/r_s^2$. In fact, a  very simple model for the dipole cross-section has been proposed by Golec-Biernat and W\"usthoff
(GBW)  \cite{GolecBiernat:1998js} with the parameterization
\begin{equation}
\sigma_{\rm dip}(x,\r_\perp)=\sigma \left[
1-e^{-\frac{1}{4}Q_s^2(x)r_\perp^2}
\right].
\end{equation}
Quite remarkably, with $Q_s^2(x)\equiv Q_0^2 \left({x_0}/{x}\right)^\lambda$,  this leads to a very good description of HERA data at $x<10^{-2}$
and moderate $Q^2$ \cite{Stasto:2000er}. The fact that the $x$ dependence of the cross section enters only in the definition of the basic scale $r_s$ has been dubbed `geometrical scaling'.

\begin{figure}[htbp]
\begin{center}
\vspace{-0.5cm}
\includegraphics[scale=0.5]{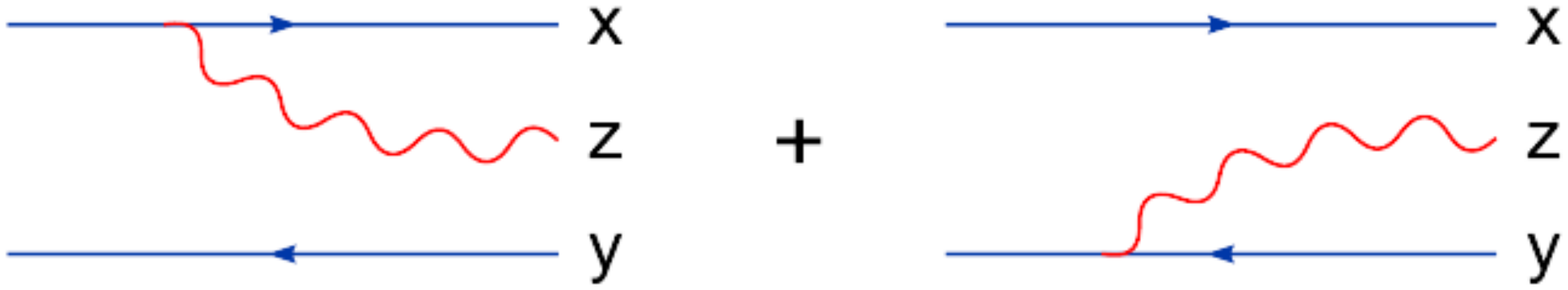}
\vspace{-7cm}
\end{center}
\caption{\label{fig:dipole_splitting} Basic effect of the evolution on the dipole. An increase of rapidity opens the phase space for the emission of one gluon. This contributes to the screening of the charge, modifying the effective interaction of the dipole as it traverses the color field of the nucleus. In the large $N_c$ limit, the dipole-gluon system is equivalent to a pair of elementary dipoles, with endpoints $\x,\z$ and $\z,\y$.}
\end{figure}

Leaving aside phenomenology, we note that $r_s$, or equivalently the saturation momentum $ Q_s$, is determined by the dynamics of the gauge fields, i.e., from QCD. In particular,  the energy dependence of $Q_s$  is expected to follow from the non  linear evolution equations that capture the phenomenon of saturation. These equations exist  in several versions. The dipole picture provides perhaps the most direct and intuitive approach to these equations. In this picture, the main agent responsible for the evolution is the emission of a gluon by a color dipole, which takes place when the dipole is boosted to a high enough rapidity. This is depicted in Fig.~\ref{fig:dipole_splitting}. After the emission of a gluon, the original dipole turns into a dipole-gluon system, whose propagation  in the field of the hadron differs from that of the original dipole. The evolution equation is precisely the equation that accounts for this modified propagation. For the correlator of two Wilson lines, this equation takes the form 
\beq
\partial_Y \langle {\rm Tr}\left(U^\dagger_{\x}U_{\y}  \right) \rangle_{_Y}=-\frac{\alpha_s}{2\pi^2}\int d^2\z\,{\cal K}_{\x \y \z}\,\langle N_c{\rm Tr}\left(U^\dagger_{\x}U_{\y}   \right)-{\rm Tr}\left(U^\dagger_{\x}U_{\z}   \right){\rm Tr}\left(U^\dagger_{\z}U_{\y}   \right)\rangle,
\eeq
where ${\cal K}_{\x \y \z}\,d^2\z \,dY$, with
\beq
{\cal K}_{\x \y \z}\equiv \frac{(\x-\y)^2}{(\x-\z)^2(\y-\z)^2}, 
\eeq
 is the coordinate space version of the branching probability (\ref{eq:branching}). 
This equation is in fact the first in an infinite hierarchy of equations for the correlators of an arbitrary number of Wilson lines that was obtained by Balitsky \cite{Balit1}.  The JIMWLK equation \cite{JIMWLK} is a functional equation that describes the evolution of the generating functional of all these correlators. Its physical content is identical to that of Balitsky's hierarchy.  
The Balitsky-Kovchegov (BK) equation \cite{Kovch3} is obtained by assuming a factorization of the non-linear term
\beq
\langle{\rm Tr}\left(U^\dagger_{\x}U_{\z}   \right){\rm Tr}\left(U^\dagger_{\z}U_{\y}   \right)\rangle\approx \langle{\rm Tr}\left(U^\dagger_{\x}U_{\z}   \right)\rangle\langle{\rm Tr}\left(U^\dagger_{\z}U_{\y}   \right)\rangle,
\eeq
which is usually justified on the basis of large $N_c$ arguments. This factorization allows to close the Balitsky hierarchy at the level of the 2-point function. 
The most  elaborate calculations to date involve solving the BK equation, with running coupling (see the contributions by Kovchegov and by Albacete). For numerical solution of JIMWLK and comparison with BK see the discussion in \cite{Weigert:2005us} and references therein.

We have presented the evolution equations as reflecting the change in the internal structure of the dipole that probe the hadron wavefunction under a boost.  Of course, moving to a frame where the dipole stays intact and the boost affects only the hadron,  we can view the emission of the extra gluon as part of the modification of the hadron wave function. Thus the evolution equations can be as well interpreted as reflecting the evolution of the hadron wavefunctions with increasing energy. 

\begin{figure}[htbp]
\begin{center}
\includegraphics[scale=0.45,angle=0]{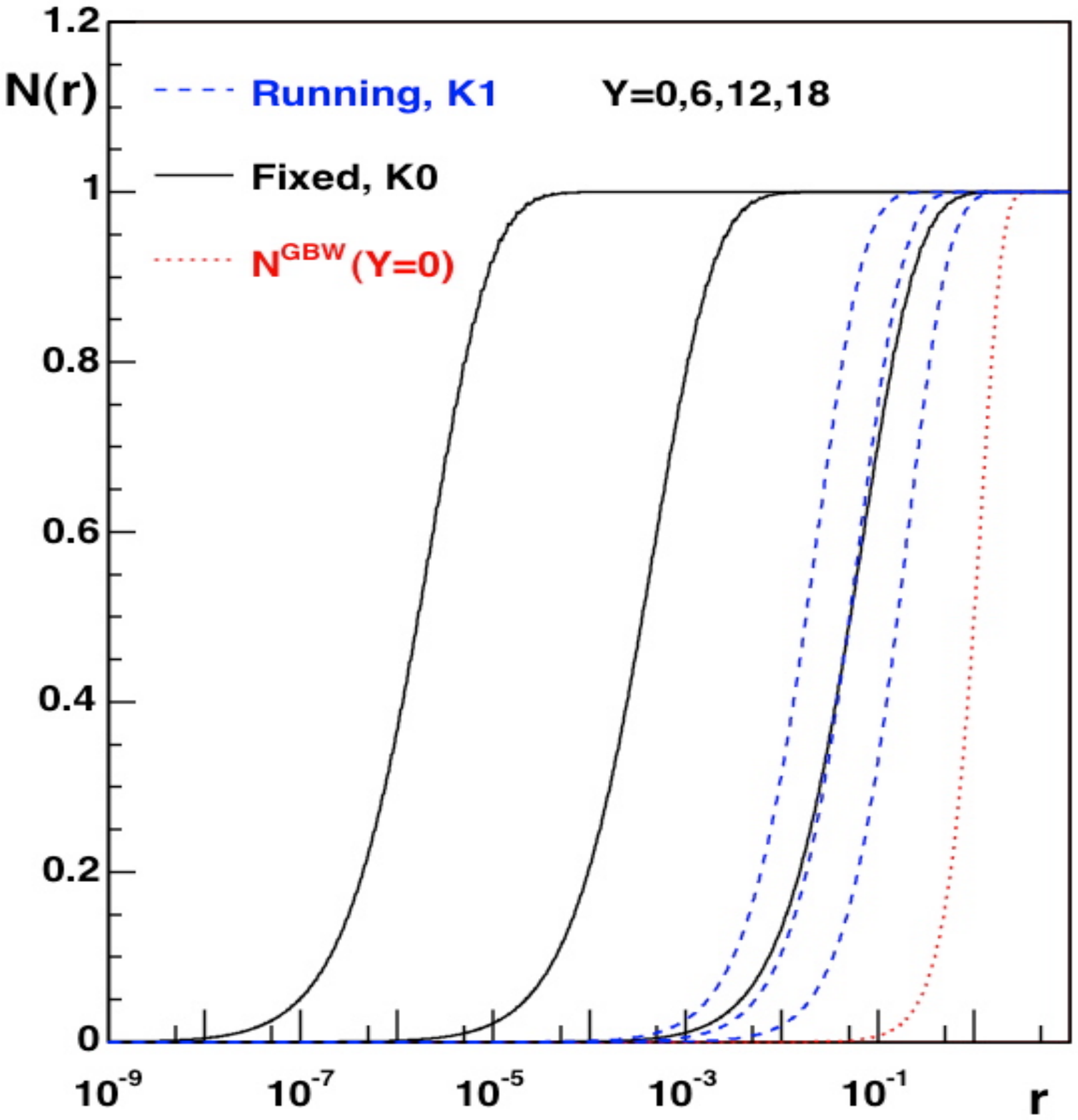}
\end{center}
\caption{\label{fig:trav} The dipole scattering amplitude $N(r)$  as a function of the dipole size for various  rapidities $Y$(from Ref.~\cite{Albacete:2004gw}).  The curves  are obtained by solving the BK equation with fixed coupling (full lines) or running coupling (dashed lines). The travelling waves are clearly visible, with a `saturation front' joining small and large dipoles at a scale $r_s(Y)$ that decreases with $Y$. Note that the evolution is slower for running coupling than for fixed coupling.}
\end{figure}

With the change of variable $N=1-S$, we can write the BK equation as an equation for the scattering amplitude $N$:
\beq
\partial_Y N_{\x \y} =-\frac{\alpha_s N_c}{\pi}\int \frac{d^2\z}{2\pi}\,{\cal K}_{\x \y \z}\,\left(    N_{\x\z}+N_{\z\y}-N_{\x\y}-N_{\x\y}N_{\z\y} \right).
\eeq
Note that the first three terms (those linear in $N$) yield the BFKL equation, in a form that underlines the structure of the hadron wave function in terms of color dipoles \cite{Mueller:1993rr}. Interesting connections have been established between this equation and equations that occur in  problems of classical statistical mechanics, e.g.  in the study of reaction-diffusion processes  (see the review by Munier \cite{Munier:2009pc}). In particular, it was shown that the asymptotic solutions of the BK equation have an universal behavior, in the form of traveling waves \cite{Munier:2003vc} that represent the evolution of the saturation front with rapidity (see Fig.~\ref{fig:trav} for an illustration) :
\beq
N(r,Y)=N(r-r_s(Y)).
\eeq
The property of geometric scaling  naturally emerges from such solutions.

\section{Classical fields, CGC}

We have seen classical fields emerging at several occasions in the preceding discussion. For instance, we often used implicitly an adiabatic approximation in which during  the time of interaction between a dipole and a hadron, the color field of the hadron can be considered as frozen, i.e., its fluctuations can be ignored. Another important feature has emerged:  near saturation the color fields are `strong', with amplitude $A\sim 1/g$, and they provide a natural description of  highly occupied, strongly coupled, modes. 

The Color Glass Condensate formalism puts these classical fields at the heart of all considerations.  Its  goal is to provide a complete description of the small $x$ component of the nuclear wave functions, thereby allowing, among other things,  the calculation of observables that control the early stages of nucleus-nucleus collisions. 

A key idea in the approach is the separation of the small $x$ degrees of freedom, treated as classical fields, from the  sources that give rise to these fields and that  are frozen during the collision.  One may imagine the various field configurations at a given rapidity $Y$ to be described by a wave-function $\Phi_Y[A] $.  Because the field does not change during the collision, the average over the field configurations will naturally involve  the square of the wave function, i.e., a probability (at least for sufficiently inclusive quantities). Schematically, the calculation of observables at some rapidity Y will then take the form
\beq
\left< \cdots \right>_Y =\int {\cal D} A \;|\Phi_Y[A] |^2 \;\langle A|\cdots |A\rangle.
\eeq
But how can we determine $|\Phi_Y[A] |^2 $?

A daring step  towards answering this question was taken by McLerran and Venugopalan. 
Their model implements the separation of scale in the most economical way, by taking the field $A_\mu$ as the solution of  the classical Yang-Mills equations:
\beq
\left[  D_\mu, F^{\mu\nu} \right] =J^\nu,
\eeq
in the presence of a frozen color source $J^\nu$. Typically, $J^\mu(x^-,x_\perp)=\delta^{\mu+}\rho(x^-,x_\perp)$ for a nucleus moving at nearly the speed of light in the positive direction. The question of determining the distribution of field configurations $|\Phi_Y[A] |^2 $ is reduced to that of finding the distribution of color charges $\rho$, traditionally denoted by $W[\rho]$. In the MV model, this is taken to be a gaussian distribution, characterized by the density-density correlation function
\beq
\langle   \rho^a(x^-,x_\perp)\rho^b(y^-,y_\perp)\rangle =\delta_{ab}\delta(x^--y^-) \delta^{(2)}(x_\perp-y_\perp) \mu^2(x^-).
\eeq
That is, one assumes that color sources are completely uncorrelated. Arguments that motivate such an ansatz can be formulated in the case of a heavy nucleus \cite{MVmodel}.

The gluon distribution provided by the MV model can be easily calculated. It provides a non-abelian generalization of the Weizs\"acker-Williams (WW)  field of electrically charged particles. This gluon distribution  is closely related  to the correlator  $S(\r_\perp)$ of two Wilson lines,  in the adjoint representation. This correlator is given by Eq.~(\ref{correlatorWL}) in which one replaces $C_F$ by $C_A=N_c$, with, accordingly 
\beq
 Q_s^2=\frac{4\pi^2\alpha_s}{N_c^2-1}n(b) xG(x,Q_s^2).
\eeq
The actual WW  gluon distribution, $\varphi_A$,  is given by 
\beq\label{MVugd}
\varphi_A(\k_\perp)=\frac{1}{\alpha_s N_c} \int d^2 \r_\perp {\rm e }^{-i\k\cdot \r_\perp}\,\frac{1- {\rm e}^{- Q_s^2 \r_\perp^2/4}}{\pi \r_\perp^2},\qquad \pi \alpha_s N_c \nabla_\k^2 \varphi_A(\k_\perp) =N(\k_\perp),
\eeq
with $N(\k_\perp)$ the Fourier transform of $N(\r_\perp)=1-S(\r_\perp)$. Note that this distribution is independent of $x$ (since each valence quark produces an $x$-independent density). 
\begin{figure}[htbp]
\includegraphics[scale=0.5,angle=00]{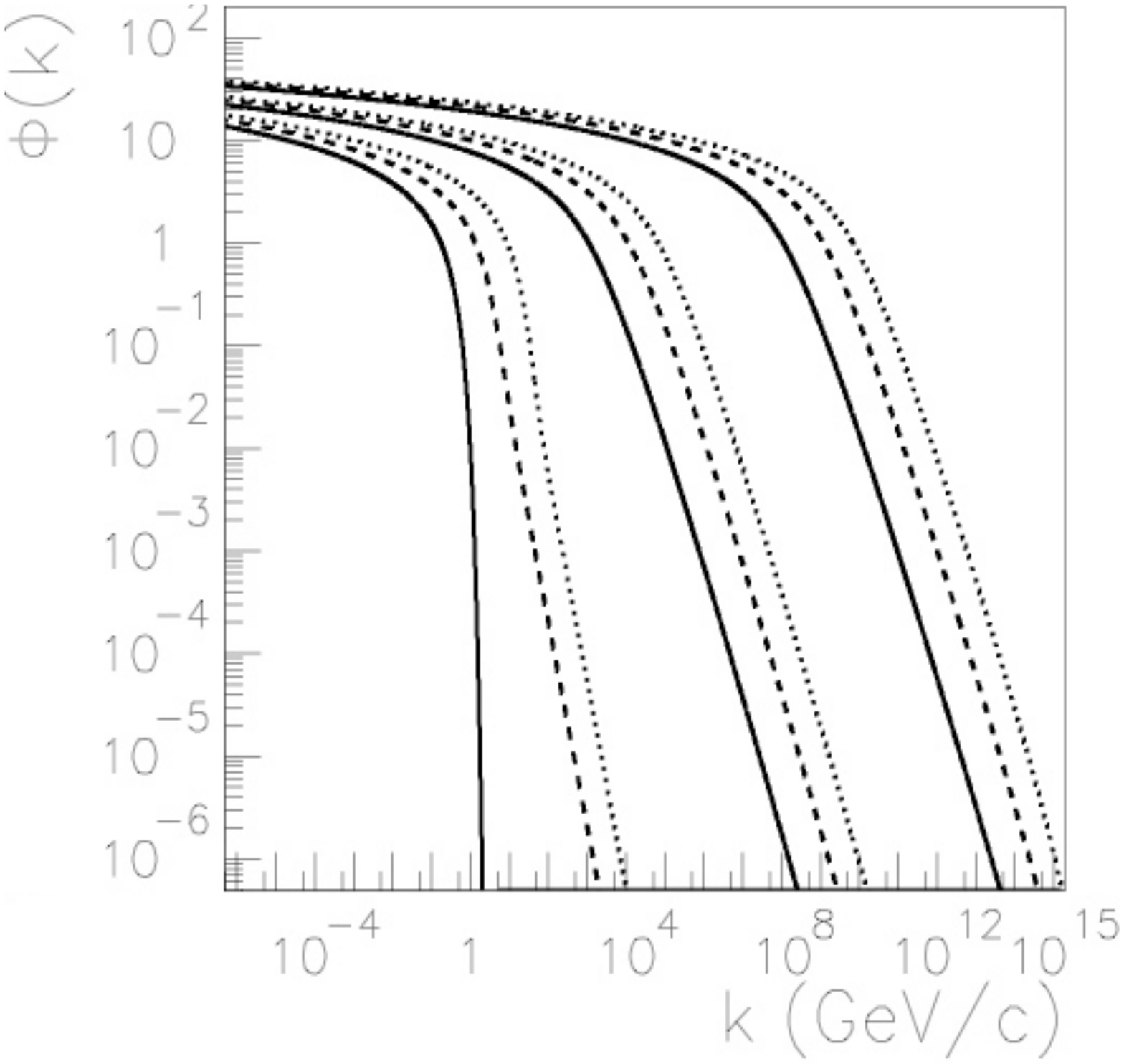}
\vspace{-1cm}
\caption{\label{fig:gluondist} The unintegrated (WW) gluon distribution obtained as solution of the BK equation (from Ref.~\cite{Albacete:2003iq}). Initial conditions are GBW (solid lines), MV with $Q_s^2=4 $GeV$^2$ (dashed lines) and MV with $Q_s^2=100$ GeV$^2$ (dotted lines). The three sets of curves correspond the rapidities $y=0, 5, 10$, respectively, from left to right. }
\end{figure}
The behaviors  for small and large transverse momenta are interesting. We have
\beq\label{MVsaturation}
\varphi_A(k_\perp\ll Q_s )\approx \frac{1}{\alpha_s N_c} \ln \frac{Q_s^2}{k_\perp^2},\qquad
\varphi_A(k_\perp\gg Q_s )\approx \frac{1}{\alpha_s N_c} \frac{Q_s^2}{k_\perp^2}.
\eeq
At small momentum, $\varphi_A(k_\perp\ll Q_s )$ exhibits saturation. It is in fact believed that the small momentum behavior  in Eq.~(\ref{MVsaturation}) is  generic, and indeed, as we shall see shortly it is essentially preserved during the evolution. It is quite remarkable that this feature emerges from the non abelian WW field calculated with uncorrelated charges.  Note that in this small momentum regime,  $\varphi_A\sim 1/\alpha_s$, reflecting the large occupation of the low momentum modes, as expected at saturation. At large momenta, the distribution decays as $Q_s^2/k_\perp^2$, in agreement with  perturbation theory and the additivity of the gluon distributions of the various sources. 
 
 The gluon distribution obtained from the MV model is often used as initial condition in the equations describing the evolution towards small $x$.  As an illustration, we show in Fig.~\ref{fig:gluondist} solutions of the BK equation that use MV  gluon distribution as initial input.   The distribution is nearly constant (in a logarithmic scale)  up to the knee that corresponds to the saturation momentum which grows with increasing rapidity. 
The evolution produces a mild increases of the height of the plateau at small $k_\perp$, but the dominant effect is the increase in the width of this plateau that reflects the increase in the saturation momentum. 
An important feature that is not made obvious by the plot concerns the modification of the power law at large momenta, which is of the form $\left(Q_s/k_\perp\right)^{2\gamma}$, with $\gamma\approx 0.63$. The numerical calculations reveal that this modification of the spectrum extends to fairly large values of the transverse momentum. This  is an intriguing feature, which seems to imply that the approach to the standard perturbative QCD behavior  is delayed  to unexpectedly  large momenta.

The WW gluon distribution is not the distribution that enters naturally the calculation of many observables.   For instance the spectrum of gluons produced in a nucleus-nucleus collisions is given by 
\beq\label{kTfactor}
\frac{dN}{dyd^2\p_\perp d^2\b}=\frac{1}{2\pi^4 C_F}\frac{\alpha_s}{p_\perp^2}\int \frac{d^2\k_\perp}{(2\pi)^2} \phi_A(x_1,k_\perp)\phi_B(x_2,|\p_\perp-\k_\perp|).
\eeq
This $k_T$ factorized expression (strictly valid only in the case where one of the two colliding systems is dilute) expresses the result in terms of an elementary cross section $\sim \alpha_s/\p_\perp^2$ and an integral over unintegrated distribution functions defined as 
\beq
\phi_A(x,\k_\perp)=\frac{k_\perp^2}{4\alpha_s N_c}\int d^2\x_\perp {\rm e}^{-i\k_\perp\cdot\x_\perp}\langle  {\rm Tr} U^\dagger (0)U(\x_\perp)\rangle _Y.
\eeq
There is a simple relation between $\phi_A(x,\k_\perp)$ and $\varphi_A(x,\k_\perp)$ (see Eqs.~(\ref{MVugd})):
\beq\label{k2phi}
\phi_A(x,\k_\perp)\sim k_\perp^2\nabla^2_k \varphi_A(x,\k_\perp).
\eeq
 Note that the total multiplicity obtained by integrating the spectrum (\ref{kTfactor}) over $\p_\perp$ is plagued with a logarithmic infrared divergence that is usually regulated by giving a small mass to the emitted gluon (or introducing a cut-off).

The assumption of uncorrelated charges of the MV model is a crude approximation: the evolution rapidly builds up correlations between the color charges. It is interesting to analyze how the various stages of the evolution modify the  density-density correlation function $\mu^2(\k)$ (see Eq.~(\ref{correlatorA+})). 
In the MV model, at the start of the evolution,  $\mu^2(\k)= \mu^2$ constant. In the BFKL regime, near saturation,  $\mu^2(\k)\sim \sqrt{\mu_A^2 \k^2} {\rm e}^{\omega\alpha_s Y}$. Finally in the deep saturation region,  
$\mu(\k)\sim \k^2 (Y-Y_s(\k))$. A  physical discussion of this evolution of the charge correlation function in terms of the gradual build up of the screening of color charges  is presented in Ref.~\cite{Mueller:2002pi}.

At this point we shall introduce the second important ingredient in the CGC formalism. It is based on the observation that the separation between charges and fields, that is exploited in the MV model,  involves a dividing scale between degrees of freedom: the field describes degrees of freedom with some particular value of $x$, while the color charge, and their correlations, are determined from degrees of freedom with $x$-values $x'>x$. As we move to lower and lower values of $x$ more and more degrees of freedom are treated as random charges (cf. the discussion of the BFKL cascade earlier), and the corresponding correlators are modified.    The requirement that observables should remain independent of the dividing scale can be implemented (to leading logarithm accuracy) as a renormalization group equation for the distribution $W_Y[\rho]$. This renormalization group equation is nothing but the JIMWLK equation. 

A further progress concerns the factorization theorem presented by Gelis. To keep the discussion concrete, let me focus on the calculation of the energy momentum tensor at the early stage of a nucleus-nucleus collision. 
In the CGC formalism, this is given by ($\tau$ is the proper time and $\eta$ the space-time rapidity)
\beq\label{Tmunu}
  \left<
    T^{\mu\nu}(\tau,{\eta},\vec\x_{\perp})
  \right>_{_{CGC}}
  =
  \int 
  \big[D{\rho_{_1}}\,D{\rho_{_2}}\big]
  \;
  { W_1\big[\rho_{_1}\big]}\;
  {W_2\big[\rho_{_2}\big]}\;
  T^{\mu\nu}_{_{\rm LO}}(\tau,\vec\x_{\perp}),
\eeq
where $T^{\mu\nu}_{_{\rm LO}}$ denote the contribution calculated from the classical fields generated by the distributions of color charges $\rho_1$ and $\rho_2$ of the respective nuclei. Equation (\ref{Tmunu}) exhibits the factorization alluded to:  the distributions $W[\rho]$ play a role analogous to the parton distributions in the collinear factorization scheme; they determine the classical field  which enters the calculation of $ T^{\mu\nu}_{_{\rm LO}}$. Because the weight functions $W[\rho]$ obey the JIMWLK equations, this formula resums all contributions of the form $(\alpha_s\ln(1/x))^n$. Besides, the dynamics of the classical field is treated exactly, which is necessary when the sources are strong, $\rho\sim 1/g$. Note that, in this approach,  
correlations, and fluctuations, are introduced through the averaging over the color charges.

\section{Phenomenology}

We are now ready to examine some of the phenomena observed at RHIC, and discuss to which extent these may reveal features of saturation discussed  in the previous sections. Two important aspects need to be kept in mind in going through this discussion. The first one is that  the phenomenology is in most cases blind to many of the details of the theory, and tests will therefore concern mainly general qualitative trends. The second aspect is that the solution of the most sophisticated evolution equation is technically difficult, and approximations are often used that tend to obscure the interpretation. 

\subsection{The basic ingredients}

Much of the phenomenology based on the saturation picture is driven by the saturation scale and its dependence on energy and impact parameter. Typically, 
\beq
Q_s^2=Q_0^2 \left(\frac{x}{x_0}  \right)^\lambda, \qquad Q_0^2(b)=Q_0^2(0) T_A(\b),
\eeq
where $T_A(\b)=\int dz \,n(\b,z)$, with  $n(\b,z)$ the nucleon density. These are of course only approximate relations, and the energy dependence of $Q_s$, for instance, can be obtained by solving an evolution equation such as the BK equation, or from the more sophisticated treatment presented in Ref.~\cite{Triantafyllopoulos:2002nz}. The treatment of the impact parameter  remains in most cases rather crude, and even the simple additive geometrical approximation suggested by the second formula above is rarely implemented. In many cases one assumes that nuclei are flat uniform cylinders,  for which $T_A(\b)=A/\pi R^2$.  

A further important ingredient is the running of the coupling constant. In the simplest treatments, one usually chooses the scale of the running coupling to be the saturation momentum itself, i.e.,  $\alpha_s=\alpha_s(Q_s)$.  But there are now sophisticated implementations of running coupling effects in the evolution equations (see the contributions by Kovchegov and by Albacete). A generic effect of the running of the coupling is to slow down the effects of the evolution.

The unintegrated gluon distribution functions that enter the calculation of most observables  have the typical shapes of the functions displayed in Fig.~\ref{fig:gluondist}. A crude approximation that  is often used consists in assuming the simple form $\varphi(k_\perp)\sim 1/\alpha_s$ for $k_\perp\ll Q_s(x)$ and
$\varphi(k_\perp)\sim Q_s^2(x)/k_\perp^2$ for $k_\perp\gg Q_s(x)$.
When multiplied by the phase space factor $k_\perp^2$ (see e.g. Eq.~(\ref{k2phi})), these unintegrated distribution functions are peaked at transverse momenta of the order of $Q_s$, so that many processes are dominated by transverse momenta of the order of $Q_s$. 

Most calculations rely on the so-called $k_T$-factorization. While this approximation can be justified for the spectrum of particles produced in the collision of a dilute system on a dense system (it exploits a linearization with respect to the color source of the dilute system), it does not hold in the case of the collision of two dense, saturated objects, like a nucleus-nucleus collision.  Recent numerical studies within the MV model show that the violation of $k_T$ factorization can be large in the region of small $k_\perp$ \cite{Blaizot:2010kh}. In practice, a cut-off is used to control this low $k_\perp$ region, and this introduces uncertainties which are difficult to quantify (although one may argue that the effect of the cut-off may vary little with energy). 

When discussing some particular situations, like for instance the physics in the  fragmentation regions, one often mixes factorization schemes, in order to deal with systems with vastly different gluon densities: one of the colliding nuclei is dense and describable by the CGC, the other is dilute, dominated by large $x$ partons, and for it collinear factorization, with standard parton distribution functions, seems appropriate. This mixing of schemes, while `reasonable', has not received a deep theoretical justification. 

An important feature of high density gluon systems is contained in various correlations. In the CGC approach, most of these correlations are generated by the average over the charge distributions. Depending on the process examined, various $n$-point functions need to be calculated.  For instance the production of quark-antiquark pairs in pA collisions involves in addition to the 2 point function, correlators of 3 and 4 Wilson lines, in the fundamental or adjoint representations \cite{Blaizot:2004wu}. The same thing happens in  the study of di-hadron production in DIS or proton-nucleus  collisions  \cite{JalilianMarian:2004da,Marquet:2007vb}. The evolution of these correlators with rapidity is usually obtained through the BK equation, that is, only 2-point correlators are evolved. This means that $n$-point correlators have to be factorized first. There are numerical indications that the full JIMWLK equation and the BK equation lead to close results  (see e.g. the discussion in \cite{Weigert:2005us}), but this may not hold for all situations.

\subsection{Deep inelastic scattering and geometrical scaling}

Before we start reviewing RHIC results, let us recall that most of what we know about parton densities comes from lepton-hadon 
DIS as measured in particular at HERA. The fits that determine these parton densities  are traditionally performed within the DGLAP evolution scheme. The state of the art was reviewed by  Forte. One important issue in the present context  is the role and importance of small $x$ effects, beyond those included in the initial conditions. Calculations at NLO and at NNLO seem capable to account for the data. However  Forte presented some evidence for tensions in  fits that may suggest a stronger small $x$ evolution than what is presently taken into account. Another issue is whether the data provide any evidence for BFKL evolution. It  was argued by Kowalski that one can  get a good fit to the data using BFKL, once NLO corrections are included,  in particular through the  $Q^2$ dependence of the power law that controls the growth of the gluon density.

A remarkable property of the data, which is beautifully captured by the GBW model, is the geometrical scaling, with  the $x$-dependence entirely hidden in the energy dependence of the saturation scale $Q_s$, as we discussed earlier. The physical origin of the scaling is still under debate.  In the dipole picture, it emerges very naturally (to within logarithmic corrections), as we have seen. It can also be understood as a universal property of the small $x$ evolution equations in their asymptotic (large rapidity) regime \cite{Iancu:2002tr}. That this property emerges also from DGLAP evolution is a priori less obvious, but it is so, as argued by Forte: not only DGLAP preserves geometrical scaling if this is built in the initial condition, but it is capable of generating scaling through evolution for not too small $Q^2$, $Q^2\gtrsim 5$ GeV$^2$. (For a recent discussion see  \cite{Caola:2008xr}.) Thus, even though geometrical scaling is very suggestive of the underlying physics of saturation, the situation remains somewhat inconclusive. 

\subsection{Particle multiplicities in hadronic collisions}

By  assuming that the bulk of produced particles comes from the saturated part of the wave functions, one immediately gets the following generic behavior for the multiplicity density
\cite{Kharzeev:2001yq}
\begin{equation}
\frac{1}{\pi R^2}\frac{dN}{dy}\sim 
\frac{Q_s^2}{\alpha_s(Q_s^2)},
\end{equation}
with all dependences on energy or centrality being entirely determined by the corresponding dependences of $Q_s$. 

Detailed calculations can of course be made beyond this crude estimate. The most sophisticated calculations  were presented by Albacete. His calculation involves solving the BK equation, with running coupling corrections. The energy dependence of the saturation scale $Q_s(x)$ is determined from the BK equation, with parameters (entering the initial conditions) obtained by fitting RHIC data (and are compatible also with HERA data). A mixed scheme is used to treat the tails of the multiplicity distributions, dominated by large $x$ partons.  Because of the poor treatment of the impact parameter, there is a large uncertainty in the overall normalization, which however depends only mildly on energy, so that the prediction for the LHC should be reasonably accurate. Another source of uncertainty is the $k_T$ factorization on which the calculation is based. 

Note that the CGC
only predicts the distribution of `initial gluons', set free typically at
a proper time $\tau\sim Q_s^{-1}$. Between this early stage and the
 freeze-out, the system undergoes several non-trivial
steps: kinetic and chemical equilibration (possibly with additional
parton production), hadronization, etc. The fact that the CGC multiplicity accounts well for the data seems to imply that there is little room left for the late stages of the collision to contribute significantly to the multiplicity. 

\subsection{Limiting fragmentation}

By  `limiting fragmentation', one refers to the property of the rapidity distribution $dN/dy$ to be 
independent of the collision energy at high energy,  in the vicinity of the beam rapidity $Y=\ln( \sqrt{s}/m)$, i.e.,  in the fragmentation region. The interval in rapidity over which this phenomenon is observed grows with energy. Evidence for such
a behavior has indeed been found at RHIC and was reviewed by W. Busza. 

In the  CGC picture one assumes that the phenomenon finds its origin in the initial gluon production, and is little affected by subsequent evolution of the system. The nucleus that sits in one fragmentation region is a dilute partonic system while the other nucleus is in
a saturated state. The rise in multiplicity when one moves away from the beam rapidity  is then attributed to the growth of the parton
distribution in the dilute nucleus.  More quantitatively, this may be seen by using the $k_T$ factorized formula (\ref{kTfactor}), recalling that the unintegrated gluon distribution functions are peaked around their respective $Q_s$. The relevant partons in the projectile (with $y\simeq Y_{beam}$) have large $x_1=({p_T}/{m}){\rm e }^{y-Y_{beam}}$, and hence small $Q_s$, while those in  the target  have small $x_2=({p_T}/{m}){\rm e }^{-y-Y_{beam}}$ and hence a large $Q_s$. It follows that the typical transverse momenta of partons in the projectile are much smaller than those of partons in the target. This remark can be exploited to perform the integration over $\p_\perp$ in Eq.~(\ref{kTfactor})) and obtain the multiplicity in the form
\beq
\frac{dN}{dy}\propto x_1G(x_1,Q^2).
\eeq
As anticipated, this depends only on $y-Y_{beam}$, and weakly on the quantities that are involved in fixing the scale $Q^2$. A detailed discussion of this phenomenon was presented by A. Stasto.

\subsection{Initial conditions for hydrodynamical evolution}

Simple properties of the initial, saturated, wave functions have interesting consequences for the initial conditions to be used in hydrodynamical calculations, as reviewed by Dumitru. The discussion is again based on the estimate of the multiplicity density using the $k_T$ factorized formula (\ref{kTfactor}).  
Since the  momentum integration is dominated by transverse momenta smaller than the smallest of the saturation momenta of the target and projectile,  the result of this integration yields the local density of produced gluons in the form\beq
\frac{dN}{d^2\s dy}\propto{\rm min}\left\{Q_{s,1}^2,Q_{s,2}^2   \right\}.
\eeq
Now each $Q_s$ depends (approximately additively) on the number of participants in the corresponding nucleus, that is 
\beq
Q_{s,1}^2\propto T_A(\s),\qquad Q_{s,2}^2 \propto T_B(\s-\b).
\eeq
This result may be contrasted to that obtained in a standard  Glauber calculation, where the local density would be simply $T_A(\s)+T_B(\s-\b)$. It follows in particular that the eccentricity predicted by the  CGC initial conditions is typically larger than that obtained with Glauber initial conditions, which affects in particular elliptic flow calculations.

\subsection{Long range rapidity correlations}

Long range rapidity correlations can only be produced at very short time. This follows from a simple causality argument recalled by Gelis. The CGC provides a natural explanation for the origin of these correlations (and an estimate for the rapidity range over which they should be expected): the color field produced at early times  (the `glasma') is arranged in coherent flux tubes of radius of order $Q_s^{-1}$, and is correlated over a rapidity interval of order $\alpha_s^{-1}$.   However,  these correlations are  not uniquely characteristic of the CGC, since most other models of particle production encompass such long range rapidity correlations. This is so, in particular, of  the string model, as recalled in Pajares' contribution. 

Long range rapidity correlations manifest themselves in a spectacular fashion in phenomena involving also strong azimuthal correlations. This is the so-called `ridge' phenomenon. 
In order to get the collimation in azimuthal angle,  collective flow seems to be required, as discussed by Gavin, Shuryak and Grassi. A noteworthy contribution was presented by Alver who introduced the concept of `triangular flow', a simple flow effect resulting from fluctuations in the initial density distribution, and which had been apparently overlooked in all previous analyses. 

Returning to correlations, let us recall that in the CGC framework, most of the  correlations arise form the averaging over the color sources.  As discussed by Lappi, when these sources are strong, of order $1/g$, a new ordering in various contributions emerges, which differs from the natural ordering in powers of the coupling constant. Interestingly, the dominant contribution leads to the negative binomial distribution, a well-known feature of multiplicity distributions of high energy hadronic collisions. The parameter $k$ which characterizes this negative binomial distribution is found to be a growing function of $Q_s$. 

\subsection{Forward rapidity}

At forward rapidity, as we have already discussed, the wave function of the projectile is  dominated by large $x$ partons, while  the target  is a dense, saturated system. One of the early indications of a rapid evolution of the wave functions (albeit mostly in the dilute system) were provided by BRAHMS data on d-AU collisions, and the rapid disappearance of the Cronin peak with increasing rapidity. Both the existence of the Cronin peak and its disappearance with increasing rapidity are natural consequences of the CGC. As shown by Albacete, BRAHMS spectra  can now be understood quantitatively from the CGC framework, using BK equation, with parameters  adjusted on DIS data.  However since other competing explanations cannot be completely ruled out (see for instance the talk by Strikman), the issue of whether the BRAHMS effect constitutes a signature of saturation remains disputable.

\begin{figure}[htbp]
   \centerline{\hfil\
     \resizebox*{!}{10cm}{\includegraphics{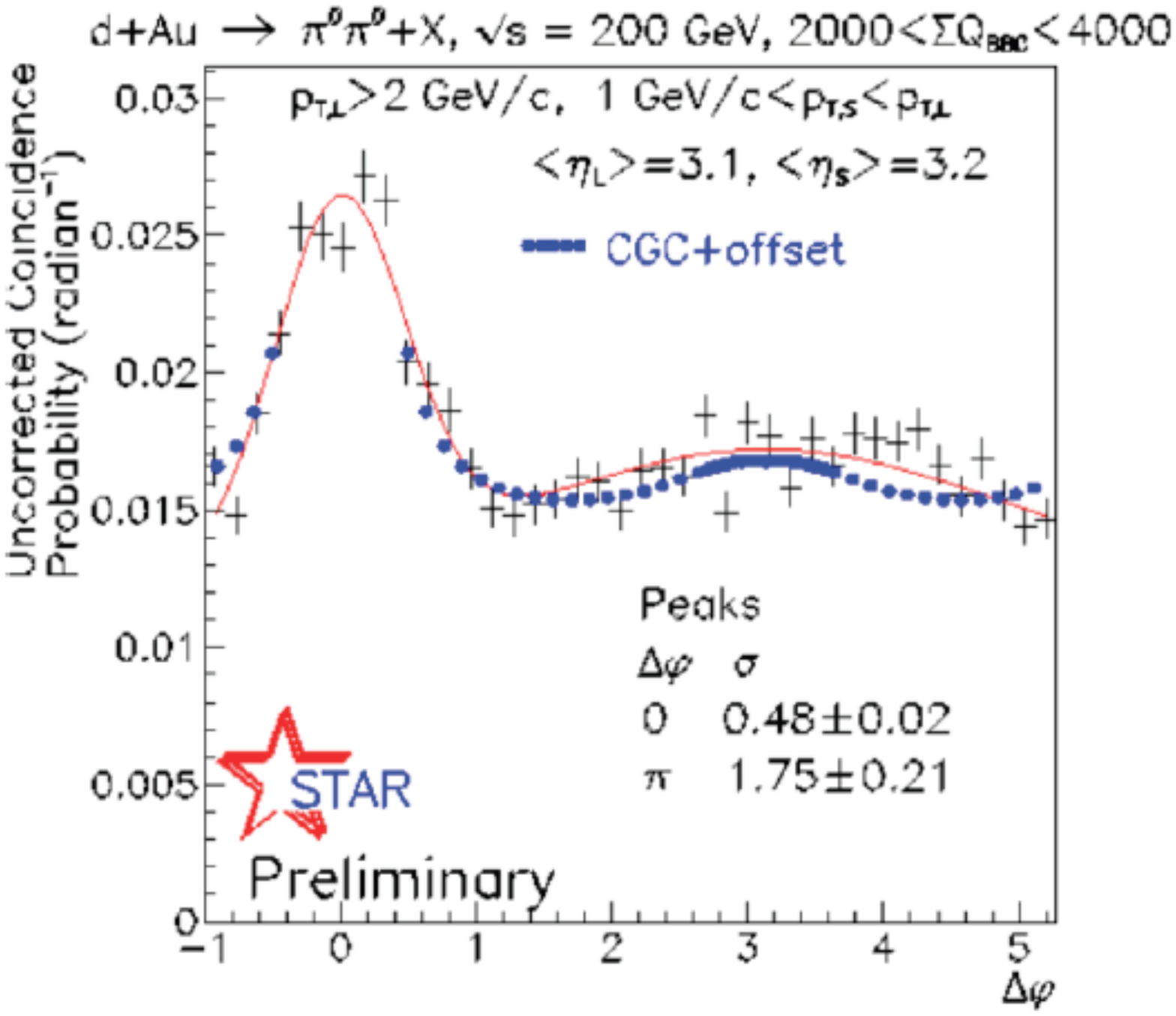}} \hfil}
\caption{\label{fig:dijet} The absence of the `away side' jet observed by the STAR experiment (taken from the contribution by Braidot). The (blue) points represent the results of the calculation by Marquet and Albacete.}
\end{figure}

Perhaps more conclusive evidence will come from 
the study of di-hadron production in d-Au collisions, as discussed by Marquet. This is indeed a very interesting situation where one can  probe the very small $x$ component of the nucleus ($x_A \sqrt{s}=k_1{\rm e}^{-y_1}+k_2 {\rm e}^{-y_2}$, with $y_1$ and $y_2$ the rapidities of the produced hadrons). The forward double inclusive pion production has been calculated  using a mixed formalism in which the wave function of the projectile is described by a standard parton distribution function while that of the nucleus is described by the CGC (using BK evolution with running coupling). The physics one expects is somewhat similar to that of the Cronin effect, namely multiple scattering in the dense gluon system. The multiple scattering of the leading quarks going through the nucleus is expected to wash out the back-to-back correlations between the produced hadrons, leading eventually to  the disappearance of the away side jet. Such an effect  has indeed been observed by STAR, and has been  reported by Braidot at this meeting (see Fig.~\ref{fig:dijet}).

There are uncertainties in the calculation that need to be clarified  before a definite conclusion can be drawn. In particular,  the calculation involves specific correlators of Wilson lines that are at the moment approximated in terms of 2-point functions (the ones that one can calculate using the BK equation). It is unknown whether the neglected higher point correlations are important or not.  But the phenomenon is very suggestive of an initial state effect that can finds its natural explanation in terms of the large gluon density of the nucleus.

\section{Summary}
 
Saturation is a generic property of QCD in the regime of high parton densities. The detailed microscopic mechanisms at work in this regime still need to be further analyzed to reach a complete understanding, but the gross features are well identified. Non linear evolution equations can be derived from QCD by following different routes, all based essentially on weak coupling approaches, the non pertubative aspects of  saturation arising from the large density of partons. (Saturation may not be limited to weak coupling, however: as shown by Iancu, an analogous phenomenon also occurs in  strongly coupled gauge plasmas, as described by the AdS/CFT correspondence.) The progress on this subject over the last decade has been truly impressive, both theoretically and experimentally. 

The progress in theory has been in fact continuous, with major steps forward at the beginning of the decade, and before,  when the non linear evolution equations were established. A better control of the saturation scale has been obtained, with more accurate determinations of its dependence on the energy and the system size.  Next to leading order corrections (in particular running coupling constant corrections) have been worked out, and when these have been implemented carefully, it has led to a better description of the data. 
In this general landscape, the  CGC (in the loose sense that it has acquired over the years),  has established itself as a reference. It has become a useful  organizing principle, suggesting new ways to look at 
the data and new measurements.  It also  appears as an essential step in building a space-time 
picture of nucleus-nucleus collisions, and this has triggered new theoretical developments.  In particular, 
new factorization theorems have been obtained, that allow for better controlled calculations of the initial stages of heavy ion collisions. The CGC relies on a separation of degrees of freedom into color charge and color fields, and on a renormalization group equation that copes with the arbitrariness in the choice of the scale at which that separation is implemented.  This confers the CGC features of an effective theory. One should however recognize that the microscopic definition of color charges, and their correlations,  still hide subtleties that need to be better understood.

The CGC, or related approaches, have led to a systematic
and successful phenomenology, based on a few basic ingredients:
the saturation momentum and its variation with energy, size 
of the systems, and simple (but not always accurate) 
approximations, such as the $k_T$-factorization.  Some features of this phenomenology 
are common to other  models (and hence not discriminant), 
but one may argue in most cases that the CGC provides a better 
connection to QCD, and the overall picture it provides 
is more systematic. Thus, although one may, rightly, argue that the data do not provide sharp and direct evidence for neither the saturation regime, nor the approach to saturation, it is fair to say that HERA and RHIC data lead to a very coherent picture, with several phenomena finding their natural interpretation in terms of high density gluonic systems. This is the case in particular of the physics at forward rapidity where the smallest $x$ components of the wave functions are being probed at RHIC energy. In fact, the beautiful results on di-hadron production that were presented at the meeting  may well constitute the first, rather direct, evidence for a large gluon density effect  in nucleus-nucleus collisons.

\vskip 3mm
\noindent{\bf Acknowledgements:} I am grateful to J. Albacete,  E. Iancu, F. Gelis, A. Mueller and J.-Y. Ollitrault for many useful discussions, and comments on the manuscript. I also wish to thank L. McLerran for his patience in waiting for this write up.  This  manuscript was partly completed during my stay as Hans. D. Jensen Professor at the Insitute for Theoretical Physics at  Heidelberg University. 



\begin{thebibliography}{100}


\bibitem{Iancu:2003xm}
  E.~Iancu, R.~Venugopalan,
  In *Hwa, R.C. (ed.) et al.: Quark gluon plasma* 249-3363.
  [hep-ph/0303204].


\bibitem{Weigert:2005us}
  H.~Weigert,
  Prog.\ Part.\ Nucl.\ Phys.\  {\bf 55}, 461 (2005)
  [arXiv:hep-ph/0501087].
  
\bibitem{Triantafyllopoulos:2005cn}
  D.~N.~Triantafyllopoulos,
  Acta Phys.\ Polon.\  B {\bf 36} (2005) 3593
  [arXiv:hep-ph/0511226].


\bibitem{JalilianMarian:2005jf}
  J.~Jalilian-Marian and Y.~V.~Kovchegov,
  Prog.\ Part.\ Nucl.\ Phys.\  {\bf 56}, 104 (2006)
  [arXiv:hep-ph/0505052].
  
 

\bibitem{Gelis:2010nm}
  F.~Gelis, E.~Iancu, J.~Jalilian-Marian and R.~Venugopalan,
  arXiv:1002.0333 [hep-ph].

\bibitem{Lappi:2010ek}
  T.~Lappi,
  arXiv:1003.1852 [hep-ph].

\bibitem{Frankfurt:2005mc}
  L.~Frankfurt, M.~Strikman and C.~Weiss,
  Ann.\ Rev.\ Nucl.\ Part.\ Sci.\  {\bf 55} (2005) 403
  [arXiv:hep-ph/0507286].
  
\bibitem{Munier:2009pc}
  S.~Munier,
  Phys.\ Rept.\  {\bf 473} (2009) 1
  [arXiv:0901.2823 [hep-ph]].
  


\bibitem{Blaizot:2002sr}
  ``QCD perspectives on hot and dense matter". Proceedings, NATO Advanced Study
  Institute, Summer School, Cargese, France, August 6-18, 2001,''  Kluwer Academic Publishers, 2002, Eds.  J.~P.~Blaizot and E.~Iancu.

  
  \bibitem{DGLAP} {G. Altarelli, G. Parisi}, Nucl. Phys. {\bf B} {\bf
    126}, 298 (1977); {V.N. Gribov, L.N. Lipatov}, Sov. J. Nucl. Phys.
  {\bf 15}, 438 (1972); {\it ibid.} 675 (1972); {Yu. Dokshitzer}, Sov.
  Phys. JETP {\bf 46}, 641 (1977).

\bibitem{BFKL} {L.N. Lipatov}, Sov. J. Nucl. Phys. {\bf 23}, 338
  (1976); {E.A. Kuraev, L.N. Lipatov, V.S. Fadin}, Sov. Phys. JETP
  {\bf 45}, 199 (1977); {I. Balitsky, L.N. Lipatov}, Sov. J. Nucl.
  Phys. {\bf 28}, 822 (1978).
  
  

\bibitem{Mueller:1994up}
  A.~H.~Mueller, 
  ``Deep inelastic scattering and small X physics,'' Lectures given at NATO Advanced Study Institute: Frontiers in Particle Physics, Cargese, France, 1-13 Aug 1994. 


\bibitem{Gribov:1984tu}
  L.~V.~Gribov, E.~M.~Levin and M.~G.~Ryskin,
  Phys.\ Rept.\  {\bf 100} (1983) 1.

\bibitem{Mueller:1985wy}
  A.~H.~Mueller and J.~w.~Qiu,
  Nucl.\ Phys.\  B {\bf 268}, 427 (1986).

\bibitem{Blaizot:1987nc}
  J.~P.~Blaizot and A.~H.~Mueller,
  Nucl.\ Phys.\  B {\bf 289}, 847 (1987).



\bibitem{Ewerz:2004vf}
  C.~Ewerz, O.~Nachtmann,
  Annals Phys.\  {\bf 322 } (2007)  1635-1669, 
  [hep-ph/0404254]; ibid. 1670-1726, 
  [hep-ph/0604087].

\bibitem{GolecBiernat:1998js}
  K.~J.~Golec-Biernat and M.~Wusthoff,
  Phys.\ Rev.\  D {\bf 59}, 014017 (1998)
  [arXiv:hep-ph/9807513]; 
  Phys.\ Rev.\  D {\bf 60}, 114023 (1999)
  [arXiv:hep-ph/9903358].

\bibitem{Stasto:2000er}
  A.~M.~Stasto, K.~J.~Golec-Biernat and J.~Kwiecinski,
  Phys.\ Rev.\ Lett.\  {\bf 86} (2001) 596
  [arXiv:hep-ph/0007192].
  
\bibitem{Balit1}
{I. Balitsky}, Nucl. Phys. {\bf B} {\bf 463}, 99 (1996).

\bibitem{Kovch3}
{Yu.V. Kovchegov}, Phys. Rev. {\bf D} {\bf 61}, 074018 (2000).

\bibitem{JIMWLK} {J. Jalilian-Marian, A. Kovner, A. Leonidov, H.
    Weigert}, Nucl. Phys. {\bf B} {\bf 504}, 415 (1997); {\it ibid.}
  Phys. Rev. {\bf D} {\bf 59}, 014014 (1999); {\it ibid.} 034007
  (1999), Erratum. {\it ibid.} 099903 (1999); {J. Jalilian-Marian, A.
    Kovner, H. Weigert}, Phys. Rev. {\bf D} {\bf 59}, 014015 (1999);
  {A. Kovner, G. Milhano, H. Weigert}, Phys. Rev. {\bf D} {\bf 62},
  114005 (2000); {H. Weigert}, Nucl. Phys. {\bf A} {\bf 703}, 823
  (2002); E. Iancu, A. Leonidov, L.D. McLerran, Nucl. Phys.
  {\bf A} {\bf 692}, 583 (2001); {\it ibid.} Phys. Lett. {\bf B} {\bf
    510}, 133 (2001); {E. Ferreiro, E. Iancu, A. Leonidov, L.D.
    McLerran}, Nucl. Phys. {\bf A} {\bf 703}, 489 (2002).

\bibitem{Mueller:1993rr}
  A.~H.~Mueller,
  Nucl.\ Phys.\  {\bf B415 } (1994)  373-385.
  
  
\bibitem{Munier:2003vc}
  S.~Munier, R.~B.~Peschanski,
  Phys.\ Rev.\ Lett.\  {\bf 91 } (2003)  232001.
  [hep-ph/0309177].
  


\bibitem{MVmodel}
 L.~D.~McLerran and R.~Venugopalan,
 Phys.\ Rev.\ D {\bf 49}, 2233 (1994);
 %
 Phys.\ Rev.\ D {\bf 49}, 3352 (1994).

\bibitem{Albacete:2004gw}
  J.~L.~Albacete, N.~Armesto, J.~G.~Milhano {\it et al.},
  Phys.\ Rev.\  {\bf D71 } (2005)  014003.
  [hep-ph/0408216].

\bibitem{Albacete:2003iq}
  J.~L.~Albacete, N.~Armesto, A.~Kovner {\it et al.},
  Phys.\ Rev.\ Lett.\  {\bf 92 } (2004)  082001.
  [hep-ph/0307179].
  
 


\bibitem{Mueller:2002pi}
  A.~H.~Mueller,
  Nucl.\ Phys.\  B {\bf 643} (2002) 501
  [arXiv:hep-ph/0206216].

\bibitem{Triantafyllopoulos:2002nz}
  D.~N.~Triantafyllopoulos,
  Nucl.\ Phys.\  {\bf B648 } (2003)  293-316.
  [hep-ph/0209121].

  
\bibitem{Blaizot:2004wu}
  J.~P.~Blaizot, F.~Gelis and R.~Venugopalan,
  Nucl.\ Phys.\  A {\bf 743} (2004) 13
  [arXiv:hep-ph/0402256]; ibid.  57
  [arXiv:hep-ph/0402257].
  
  
\bibitem{JalilianMarian:2004da}
  J.~Jalilian-Marian, Y.~V.~Kovchegov,
  Phys.\ Rev.\  {\bf D70 } (2004)  114017.
  [hep-ph/0405266].

   
\bibitem{Marquet:2007vb}
  C.~Marquet,
  Nucl.\ Phys.\  A {\bf 796} (2007) 41
  [arXiv:0708.0231 [hep-ph]].

  

\bibitem{Iancu:2002tr}
  E.~Iancu, K.~Itakura, L.~McLerran,
  Nucl.\ Phys.\  {\bf A708 } (2002)  327-352.
  [hep-ph/0203137]; 
  A.~H.~Mueller, D.~N.~Triantafyllopoulos,
  Nucl.\ Phys.\  {\bf B640 } (2002)  331-350.
  [hep-ph/0205167].

  

\bibitem{Caola:2008xr}
  F.~Caola and S.~Forte,
  Phys.\ Rev.\ Lett.\  {\bf 101} (2008) 022001
  [arXiv:0802.1878 [hep-ph]].


\bibitem{Kharzeev:2001yq}
  D.~Kharzeev, E.~Levin and M.~Nardi,
  Phys.\ Rev.\  C {\bf 71} (2005) 054903
  [arXiv:hep-ph/0111315].


\bibitem{Albacete:2007sm}
  J.~L.~Albacete,
  Phys.\ Rev.\ Lett.\  {\bf 99} (2007) 262301
  [arXiv:0707.2545 [hep-ph]].


\bibitem{Blaizot:2010kh}
  J.~P.~Blaizot, T.~Lappi and Y.~Mehtar-Tani,
  Nucl.\ Phys.\  A {\bf 846}, 63 (2010)
  [arXiv:1005.0955 [hep-ph]].

\end{thebibliography}
\end{document}